\documentclass[aps,pra,twocolumn]{revtex4-1}

\pdfoutput=1
\usepackage{graphicx}
\usepackage{amsmath}
\usepackage{amsfonts}
\usepackage{amssymb}
\usepackage{natbib}
\begin{document}

\title{Rydberg atom formation in strongly correlated ultracold plasmas}
\author{G. Bannasch}
\email[]{bannasch@pks.mpg.de}
\author{T. Pohl}
\affiliation{Max Planck Institute for the Physics of Complex Systems, N\"othnitzer Str.
38, D-01187 Dresden, Germany}
\date{\today}
 
\begin{abstract}
In plasmas at very low temperatures formation of neutral atoms is dominated by collisional 
three-body recombination, owing to the strong $\sim T^{-9/2}$ scaling of the corresponding recombination
rate with the electron temperature $T$. While this law is well established at high 
temperatures, the unphysical divergence as $T\rightarrow0$ clearly suggest a breakdown in
the low-temperature regime. 
Here, we present a combined molecular dynamics-Monte-Carlo study of electron-ion
recombination
over a wide range of temperatures and densities. Our results reproduce the known behavior of the 
recombination rate at high temperatures, but reveal significant deviations with decreasing temperature.
We discuss the fate of the kinetic bottleneck and resolve the divergence-problem as the plasma enters 
the ultracold, strongly coupled domain.
\end{abstract}

\maketitle
\section{Introduction}
\label{sec_introduction}
Since the first creation of ultracold plasmas (UCPs) via photo-ionization of laser-cooled
atoms
\citep{killian1999unpCreation} or cold molecules \citep{grant2008supersonic}, these
systems prove to provide a well suited platform to study a range
of plasma physics phenomena, such as collective waves
\cite{bes03,fzr06,mlt09,cmk10,lpr10,mes11,shu11b}, plasma 
expansion into vacuum \cite{kkb00,roh02,roh03,ppr03,ppr04,lgs07,tro10, grant2009nobeam},
plasma instabilities \cite{zfr08,ros11} and recombination of neutral atoms
\cite{killian2001RydFormationInPlasma,gls07,fletcher2007tbr_temp,pohl2008rates,br08}.
Besides opening up a new parameter regime \cite{killian2007UNPreview,Killian07,kir10} as
well as promising applications for nanotechnology \cite{cgt05,hcd06,rkt09,css11}, UCPs
offer a rare opportunity to study strongly correlated plasmas
\cite{ich82,red97,don99,bhb10} in the laboratory.

The degree of correlations is characterized by the Coulomb coupling parameter
$\Gamma = \frac{e^2}{a \, k_\text{B} T}$, where $e$ is the electron charge, $k_\text{B}$
is the Boltzmann constant, and
$a = (\frac{4}{3}\pi\rho)^{-\frac{1}{3}}$ is the Wigner-Seitz radius for a plasma of
density $\rho$.
When the average potential energy $\sim e^2/a$ of the charges exceeds their thermal energy
$\sim k_\text{B}T$, i. e. when $\Gamma >
1$, the plasma is termed strongly coupled. Strong coupling phenomena studied in UCPs
include disorder-induced heating \cite{mur01,mck02,kun02,scg04,cdd05,bdl11} accompanied by
kinetic energy oscillations \cite{zwicknagel1999relaxation,csl04,pohl2005relaxation} as
well as liquid-like \cite{shu11} and crystalline \cite{pohl2004crystallization,bff05,kbo10} behavior
of the ionic plasma component. 

However, in contrast to dense strongly coupled plasmas UCPs have a rather short 
life time on the order of several $10^2\mu$s.
The dominant decay mechanism is collisional recombination leading to the formation of
highly excited neutral Rydberg atoms. The corresponding rate constant $\nu\sim T^{-9/2}$
\citep{mansbach1969} has a strong dependence on the electron temperature $T$ and can,
hence, assume large values in UCPs. In the underlying three-body recombination (TBR)
process two electrons collide in the vicinity of an ion to form a weakly bound
Rydberg atom, where the remaining electron carries away the corresponding excess energy.
Subsequent collisions of the formed Rydberg atom with free electrons can further de-excite
the atom, and the released binding energy leads to an increase of the free electron
temperature $T$ in the course of the plasma evolution \cite{roh02,roh03,ppr04,fzl07,gls07}.

The strong temperature dependence of the recombination rate can be readily established
from simple scaling arguments.
The collision frequency of an electron with average velocity $\bar{v} \sim \sqrt{T}$ is
given by $\nu_c
= \bar{v} \, b^2 \, \rho_e$, where $\rho_e$ is the electron density and $b \sim T^{-1}$
is the characteristic distance of closest approach.
Since the probability of finding another electron within a distance $b$ of the colliding
pair is $b^3 \, \rho_e$, one obtains a TBR rate of $\nu \sim \nu_c \, b^3 \rho_e \sim
\rho_e^2 \,
T^{-\frac{9}{2}}$. In their seminal paper \cite{mansbach1969} Mansbach and Keck presented
a detailed study of TBR in ideal plasmas.
Performing classical trajectory Monte-Carlo (CTMC) simulations of isolated three-body
collisions and using rate equations
they confirmed the  $T^{-\frac{9}{2}}$-scaling of the recombination rate and calculated
the corresponding proportionality constant.
Subsequent studies based on CTMC calculations or rate equations
\cite{stevefelt1975rates,vs80,zygelman2005,pohl2008rates} have since confirmed the strong
temperature dependence,  which was found to be in good agreement with experiments in hot
and cold ($T>10000$ K) plasmas \cite{hh62,stevefelt1975rates,vs80} as well as with
measurements on UCPs with moderate coupling strength ($\Gamma \lesssim 0.2$)
\citep{fletcher2007tbr_temp,pohl2008rates}.

On the other hand, the strong temperature dependence of the TBR rate ultimately suggests
unphysically fast recombination in the ultracold regime. It was one of the major
motivations of early UCP experiments \citep{killian1999unpCreation,
killian2001RydFormationInPlasma} to shed light on this apparent divergence problem of the
recombination rate. 
The conflicting timescales and the role of particle correlations become particularly
evident by transforming to dimensionless units, employing the electronic plasma frequency
$\omega_\text{p} = \sqrt{4 \pi e^2 \rho_e/m}$ ($m$ is the electron mass)
such that
\begin{equation}
      \nu \sim \rho_e^2 \, T^{-9/2} \sim \omega_\text{p} \, \Gamma^{9/2}.
      \label{tbr}
\end{equation}
Consequently, for $\Gamma\gtrsim1$ the recombination rate $\nu$ would become comparable 
or larger than the plasma frequency $\omega_\text{p}$, whose inverse determines the
typical timescale of
electronic motion in the plasma. The unphysical crossover of timescales at
$\Gamma\gtrsim1$ indicates  a breakdown of the conventional recombination theory in terms of
three-body collisions and suggests a modification of the recombination rate in the
strongly coupled regime.

This fundamental problem has been addressed theoretically in a number of previous
articles \cite{bs93a,bs93b,ksb93,bos94,bonitz2000recombination,hahn1997density_recrate,
hahn01,
zygelman2003,zygelman2005,norman2009recombination}. In
\cite{bs93a,bs93b,ksb93,bos94,bonitz2000recombination} strong coupling in a dense plasma,
where quantum 
effects become important, has been considered and an enhancement of the recombination 
rate coefficient (and ionization rate coefficient) due to correlation-induced continuum
lowering has been found. On 
the other hand, studies of recombination in classical plasmas based on analytical
estimates or 
numerical calculations
\cite{hahn1997density_recrate,hahn01,norman2009recombination,bbz11} have found a
suppression of recombination in the moderately to strongly coupled regime, but the
proposed modifications of the recombination rate yield different and contradictory results
for the temperature scaling,
such that the question of Rydberg atom formation in correlated plasmas still remains an
unsettled issue.

In this article, we investigate the classical recombination of a single ion immersed in a
strongly coupled electron plasma without employing any additional approximation. Our
calculations are based
on Monte-Carlo (MC) sampling of classical molecular dynamics (MD) simulations that
provide a 
natural extension of previous three-body CTMC calculations
\cite{mansbach1969,pohl2008rates} to account for strong 
electron-electron correlations and many-body interactions. Our results quantitatively
reproduce 
the known behavior of the recombination rate \cite{pohl2008rates} for $\Gamma\ll1$ and are
consistent with the results
of Kuzmin and O'Neil \cite{kuzmin2002numsimplasma} for the particular value of
$\Gamma = 0.6$ studied in that work. We further
discuss simulations of two-component plasmas for various initial configurations and
temperatures, 
that demonstrate strong disorder-induced electron heating to $\Gamma\sim0.5$. This
conclusively 
excludes the possibility of a metastable plasma state with orders-of-magnitude suppression
of 
recombination as suggested in
\cite{norman2009recombination,bbz11}. Nevertheless, strong
coupling effects on the recombination 
rate are found to have observable consequences during the short-time evolution of UCPs.

The article is organized as follows. Details of the plasma model as well as the numerical
approach 
are given in section \ref{sec_model}, where we also review the rate equation description
of TBR 
based on CTMC collision rates. In section \ref{sec_bottleneck} we discuss the obtained
behavior of the 
bottleneck binding energy as function of $\Gamma$, allowing to calculate the recombination
rate, presented
in section \ref{sec_rate}. Finally, section \ref{sec_full} provides simulation results for
two-component neutral plasmas and a discussion of competing heating effects and their
consequences for the recombination dynamics.

\section{Numerical approaches}
\label{sec_model}
In order to study Rydberg atom formation at a constant temperature and to isolate the
recombination 
from other collision processes (see section \ref{sec_full}) we first consider the
recombination dynamics
of a single ion placed inside an electronic one component plasma (OCP), consisting of $N$
electrons and a 
homogeneous neutralizing positively charged background. While this model can only provide
a simplified 
description of a neutral two-component system (see section \ref{sec_full}) it resembles
the situation 
of antihydrogen production experiments carried out at CERN
\cite{amoretti2002antihNature,gabrielse2002antih,gab05}.
Here highly excited antihydrogen Rydberg atoms are formed via successive transits of
antiprotons through an 
ultracold positron plasma. In these experiments atoms are formed predominantly via
collisional recombination, 
which is however considerably modified by the presence of applied strong magnetic fields
as has been extensively 
studied via CTMC calculations \cite{gln91,rha04,psg06,bdu09,rob08,pss09}.

Our simulations proceed in three steps. First, we equilibrate the electron OCP
to a
predefined temperature 
$\Gamma^{-1}$. Subsequently, we place a neutral atom at the center of the cubic
simulation box, 
consisting of an ion and an electron at the origin with the electron having an excess
kinetic energy of 
$\frac{3}{2 \Gamma} \, \frac{e^2}{a}$. This procedure ensures that the total potential
energy is not affected by the introduction 
of the additional ion and at the same time gives a good description of atomic
single-photon ionization used 
to produce UCPs \cite{killian1999unpCreation}. Following the escape of the
"photo-electron" we monitor the evolution of the surrounding plasma electrons.

\subsection{Molecular Dynamics Simulation}
\label{sec_md}
The computationally most demanding part of the simulation is the evaluation of the mutual
interactions 
between the $N$ plasma electrons. Due the $\mathcal{O}(N^2)$ of the corresponding
numerical effort 
and the necessity to implement periodic boundary conditions (PBC) a straightforward force
calculation would
be prohibitively demanding in view of the accuracy and ensemble size required for the
present study.

Both of these problems can be efficiently resolved by the fast multipole method (FMM)
\cite{Greengard1987FMM} which permits force calculations for large particle numbers with a
complexity of $\mathcal{O}(N)$.
The FMM algorithm divides the simulation volume into a hierarchy of cubic subcells, and
determines multipole expansions for the charge distribution in each cell.
The interaction of a certain particle with the particles in a distant cell can then be
calculated 
much more efficiently and with a controllable error. Moreover, the FMM provides a natural
implementation 
of PBC, as the periodic images of the simulation box can be treated as an upper
extension of the hierarchy of cubic subcells \citep{lambert1996periodicFMM}. The
particular implementation of the FMM used in this work  is described in
\citep{fmm_article, fmm}.
As discussed below, our algorithm also requires to calculate interactions among a
small subgroup of particles. In this case, we perform a direct force summation and
implement PBC via standard Ewald summation \citep{ewald1921,ewald1996survey}.

In order to initiate the photo-electron at the central ion position we remove the
singularity 
of the attractive electron-ion potential according to

\begin{equation}
  V_\text{ion}(r) = \begin{cases}
          -\frac{1}{r}      & r \ge r_{\rm c}\\
          \frac{3}{2r_{\rm c}}\left(1-\frac{r^2}{3r_{\rm c}^2}\right) & r \le r_{\rm c}.
        \end{cases}
\end{equation}
where the softcore radius $r_{\rm c}=10^{-2} a$ was chosen sufficiently small to
have no influence on the simulation results, as has been checked by varying the value of
$r_{\rm c}$.

\begin{figure}[t!]
 \includegraphics[width=0.6\columnwidth]{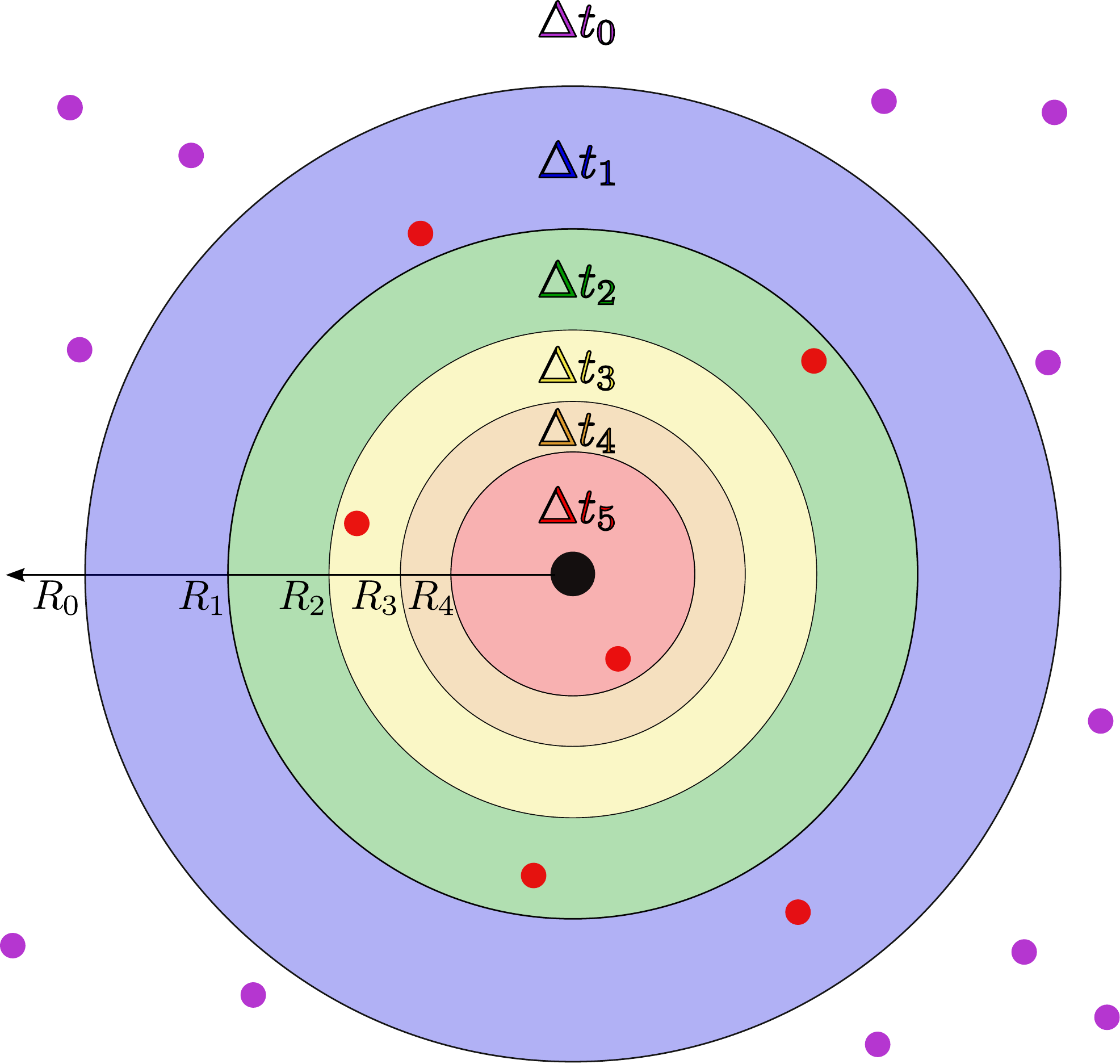}
 \caption{(color online) Two dimensional projection of the center of the simulation cell
          with central ion (black dot) and the regions of different time steps.
          All electrons within the sphere $r < R_0$ (red dots) are propagated with the
          same reduced time step $\Delta t_n$ which is determined by the smallest
          electron - ion separation $r_\text{min}$, according to $R_n < r_\text{min} <
          R_{n-1}$ (this example: $\Delta t_n = \Delta t_5$).
          Electrons with $r > R_0$ (blue dots) are propagated with the global time step
          $\Delta t_0$.
          }
 \label{radii}
\end{figure}

Consequently, the magnitude of the electron-ion force is limited by $F^{ei}_\text{max} =
1/r_{\rm c}^2$.
Typically, $F^{ei}_\text{max}$ will be much larger than the characteristic force between
the electrons
$F^{ee} \sim \Gamma^{-2}$, which induces two vastly disparate timescales of the electron
motion:
Far away from the central ion electrons move comparatively slowly on a timescale
$\sim\omega_{\rm p}^{-1}$, 
while close to the ion scattered and bound electrons undergo a much faster dynamics.
An efficient symplectic propagation of the electrons that exploits different system time
scales can be realized by the so-called reversible reference system propagator
algorithm (r-RESPA) \citep{RESPA1992, RESPA1996}.
In the present case, the principle idea is to evaluate the dynamics of distant electrons
(blue dots in Fig. \ref{radii}) with a fixed, coarse-grained time step, while the electron
motion in the immediate ion vicinity
is resolved with a smaller, dynamically adapted time step.
This guarantees an accurate description of the recombination process while maintaining
computational costs at a minimum.

Usually the r-RESPA split is either based on forces or on particles \citep{RESPA1996}.
The two time scales discussed above call for a force based r-RESPA split by separating
terms involving electron - ion interactions from terms which involve electron - electron
interactions.
However, as the condition $F^{ei}_\text{max} \gg F^{ee}$ holds only for those few
electrons which are close to the ion, it is more efficient to additionally split the
Hamiltonian
 based on particles, by separating terms involving electrons with positions $r_i < R_0 =
a$ from terms 
involving more distant electrons with $r_i > R_0$.
The total Hamiltonian $\mathcal{H}$, describing the dynamics of the $N$ electrons with
position $\mathbf{r}_i$ and momentum $\mathbf{p}_i$, ($i = 1 \dots N)$, is thus split
according to
\begin{figure}[t!]
 \includegraphics[width=0.6\columnwidth]{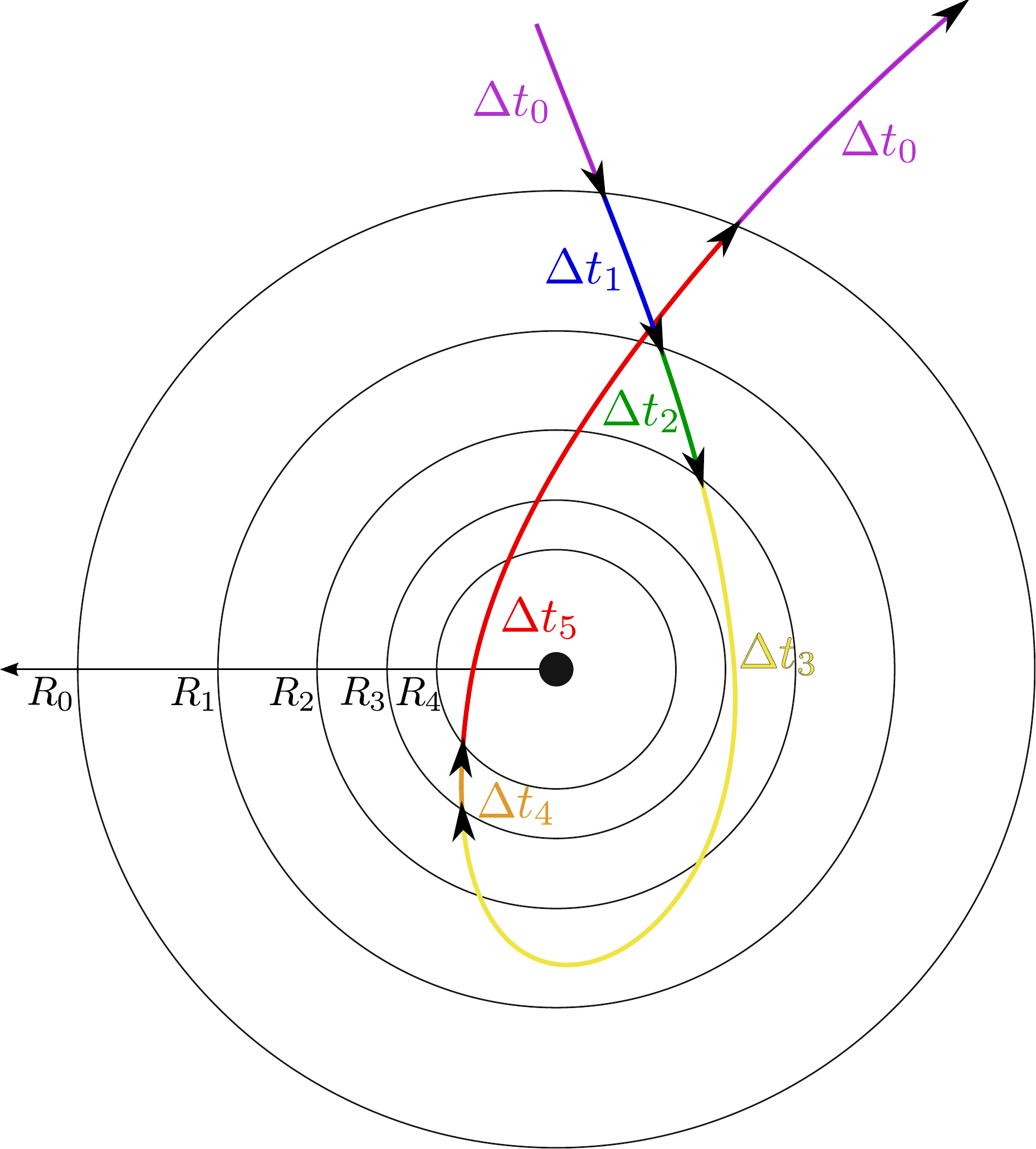}
 \caption{(color online) Schematics of the time step adaption as described in the text.}
 \label{timestep}
\end{figure}
\begin{equation}
  \mathcal{H} = K^f + K^s + V^f + V^s + E_{\text{BG}}\;,
  \label{hamiltonian}
\end{equation}
where
\begin{align}
  K^f & = \sum\limits_{i = 0 \atop r_i < R_0}^N \frac{\mathbf{p}_i^2}{2 \, m}, \notag \\
  V^f & = \sum\limits_{j>i \atop r_i, r_j < R_0}^N
    \frac{e^2}{\left|\mathbf{r}_i - \mathbf{r}_j\right|} +
    \sum\limits_{i = 0 \atop r_i < R_0}^N V_\text{ion}(r_i) \notag
\end{align}
denote the kinetic energy and interactions of particles in the region of fast dynamics
($r < R_0$).
Likewise, the remaining energies in the region of slow dynamics ($r>R_0$) are given by
\begin{align}
\label{eq_slow}
K^s & = \sum\limits_{i = 0 \atop r_i > R_0}^N \frac{\mathbf{p}_i^2}{2 \, m}, \notag \\
V^s & = \sum\limits_{j>i}^N \frac{e^2}{\left|\mathbf{r}_i - \mathbf{r}_j\right|} +
          \sum_{\mathbf{L} \ne \mathbf{0}} \sum\limits_{j,i}^N
          \frac{e^2}{\left|\mathbf{r}_i - \mathbf{r}_j + \mathbf{L} \right|} + \notag \\
    & + \sum_\mathbf{L} \sum\limits_{i = 0}^N
      V_\text{ion}(\left|\mathbf{r}_i+\mathbf{L}\right|) -
          V^f 
\end{align}
and $E_\text{BG}$ is the particle - background interaction energy.
The sum $\sum_{\mathbf{L}}$ runs over all possible lattice vectors in the periodic
lattice of image simulation boxes.
Because of TBR collisions, which will take place in the region of fast dynamics, the
electron - electron interaction among electrons in this region can become comparable to
the ion - electron interaction and has thus been included in $V^f$.

The majority of electrons participate in the slow dynamics and are propagated
with a fixed global time step $\Delta t_0$.
The corresponding electron-electron forces including image charges are calculated
within the FMM, while
the electron-ion interaction is obtained by direct force summation combined with the Ewald
summation to implement the PBC. Note that the forces arising from the image charges change
only slowly and, hence, can be entirely accounted for in the slow dynamics (cf.
eq. \ref{eq_slow})

Each electron within the critical distance $R_0 = a$ participates in the fast dynamics.
For appropriate time step adaption we divide this spherical region into concentric shells
of 
radii $R_n =2^{-n/2} R_0$, $n=1,2,...$ (see Fig. \ref{radii}). Once the smallest
electron-ion 
distance falls below $R_n$ the time step for all electrons within $R_0$ is decreased 
to $\Delta t_{n+1} =2^{-(n+1)} \Delta t_0$ (see Fig. \ref{radii}).
For the electron propagation in the region of fast dynamics with the small
time step $\Delta t_n$ the interaction $V^f$ is calculated via direct force summation and
the Ewald potential
\citep{ewald1921,ewald1996survey,brush1966OCPewald,hansen1973OCPewald,slattery1980OCP}.
In order to avoid periodic changes of the time step $\Delta t_n$, the time step is only 
decreased as long as there are electrons in the region $r < R_0$. Only when an electron
leaves
the region of fast dynamics its propagation is switched back to the global time step
$\Delta t_0$ (see Fig. \ref{timestep}).
We set $\Delta t_0 = 10^{-3} \, \omega_\text{p}^{-1}$, ensuring accurate propagation for
both free and bound electrons. In fact, we achieve a very small maximum relative energy
error in all simulations  below $0.01 \%$. We find that such a high accuracy is needed in order to obtain
reliable converged results for the recombination rate.

Formally, our propagator for the fast electrons can be written as
\begin{equation}
  S^f(\Delta t_n) = U_{V^f}(\frac{\Delta t_n}{2}) \, U_{K^f}(\Delta
    t_n) \, U_{V^f}(\frac{\Delta t_n}{2}), 
\end{equation}
where we have used the notation employing propagators
$U_A(t) = e^{t \, \mathcal{D}_A}$
of differential operators
$\mathcal{D}_A = \{\mathbf{q}, A\}$.
Here, $\{\dots\}$ denotes the Poisson bracket, $\mathbf{q} = (\mathbf{r_1}, \mathbf{r_2}
\dots \mathbf{r_N}; \mathbf{p_1}, \mathbf{p_2}
\dots \mathbf{p_N})$, and $A$ stands for any of the energy terms of the system Hamiltonian
\eqref{hamiltonian}.
The total system propagator is then given by
\begin{equation}
   S(\Delta t_0) = U_{V^s}(\frac{\Delta t_0}{2}) \, U_{K^s}(\Delta t_0) \,
    \left( S^f(\Delta t_n) \right)^{2^n}
    U_{V^s}(\frac{\Delta t_0}{2}).
\end{equation}

Besides the numerical challenges, a proper analysis of the MD data and in particular the
identification and characterization of the formed classical bound states poses additional 
difficulties. The most straightforward way would be to monitor the electrons total energy
$E_i$,
i.e. the sum of the $i$-th electron's kinetic energy and potential energy due to the ion,
the remaining $N-1$ electrons, their periodic images and the positive homogeneous
background, and declare an
electron bound when $E_i<0$. While this criterion makes sense for isolated atoms, in the 
present case many-body interactions and possibly strong electron-electron correlations 
lead to a lowered and fluctuating ionization threshold \cite{mw98}. In fact, already for
moderate coupling strength one finds that $E_i<0$ for several electrons.
Those electrons can be considered as weakly localized in slowly fluctuating potential 
wells formed by the correlated charges \cite{dkg02}, but they are not bound by 
the ionic potential. Hence, once an electron becomes bound to the ion, its energy should 
drop significantly below the energies of free electrons. Therefore, we use the lowest
electron 
energy $E_\text{min}(t) = \min \limits_{i \in N}(E_i(t))$ as an initial criterion to
select a possibly 
bound electron, whose index is denoted by $j_\text{min}(t)$.

\begin{figure}[t!]
 \includegraphics[width=\columnwidth]{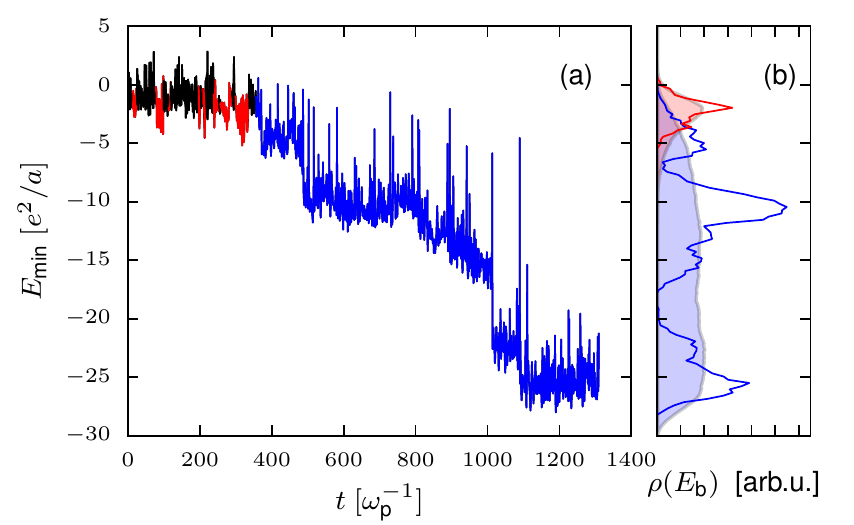}
 \caption{(color online) (a) Typical time evolution of the minimum electron energy
          $E_\text{min}$. Red parts correspond to bound
          electrons that are subsequently ionized, blue parts correspond to bound
          electrons that reach $E_\text{sink}$ without ionization for $\phi_{\rm c} = 4
          \pi$.
        (b) Corresponding energy densities $\rho_{\text{ion}}$ and $\rho_{\text{rec}}$ of
          the example trajectory (lines) and as obtained from the ensemble average
          (shaded).}
 \label{energy}
\end{figure}

A typical example of such an minimal energy trajectory $E_{\text{min}}(t)$ is shown in 
Fig. \ref{energy}a. As can be seen, $E_{\text{min}}$ stays negative most of the time,
and therefore provides only a necessary but not sufficient criterion for bound state
assignment.
In order to detect stable electron-ion orbits we adopt the procedure proposed in
\cite{gsr07} and
integrate the rotation angle $\phi$ of the $j_\text{min}$th electron around the ion.
If $j_\text{min}$ changes its value, e.g. due to an exchange collision, integration starts
again at
zero.
If the maximum rotation angle $\phi_{max}(j_\text{min})$ of the $j_\text{min}$th electron
exceeds 
a critical angle $\phi_{\rm c}$ the electron is considered to be bound.
The binding energy $E_\text{b}(t)$ of a Rydberg atom corresponds to the parts in
$E_\text{min}(t)$ where $j_\text{min}(t)$ fulfills this angular criterion:
\begin{equation}
 E_\text{b}(t) = \begin{cases}
            0           & \phi_{max}(j_\text{min}(t))  <  \phi_{\rm c} \\
            E_\text{min}(t)  & \phi_{max}(j_\text{min}(t)) \ge \phi_{\rm c},
          \end{cases}
\end{equation}
i.e. $E_\text{b}(t) = 0$ in the absence of a bound state, as marked by the black segments
in Fig. \ref{energy}a. The influence of the precise value of $\phi_{\rm c}$ on the
extracted recombination dynamics will be discussed below.

Following the initial capture, subsequent electron-atom collisions slightly (de)-excite
the formed atom, lead to re-ionization or occasionally drive the atom to significantly
deeper binding energies. Such close collisions typically cause electron exchange
\cite{gln91} and are marked by sharp peaks in $E_{\rm b}(t)$ (see Fig. \ref{energy}a).

The simulation is stopped as soon as $E_\text{b}(t)$ reaches a certain energy sink
$E_\text{sink}$. The energy sink was set to $E_{\text{sink}} / (e^2/a) = -20$ for $\Gamma
\ge 1$ and to $E_{\text{sink}} / (e^2/a) = -50$ for $\Gamma < 1$ to ensure that the
probability for a re-ionization vanishes well before the energy sink is reached (see
Fig. \ref{ionprob}) and therefore does not affect the final results for the determined
recombination dynamics.
%This implies that the the lower binding energy cut off $E_\text{sink}$ is well below the
% kinetic bottleneck (see section \ref{sec_bottleneck}) such that a further shift of the
% sink to even lower binding energies has no influence on the results of our calculations.

Physical quantities are extracted from averages over an ensemble of
$\sim10^3$ simulation runs per
parameter set, produced in a Monte-Carlo sampling over the initial positions and
velocities of
the electrons.

\subsection{Rate Equations}
\label{sec_rateequations}
To make direct comparison to previous CTMC calculations, we also solved the corresponding
rate equations for the recombination scenario discussed above.
Here one calculates the evolution of level population densities $\rho_n$ of the
recombining atom 
\citep{bates1962rates, burgess1976rateequations, stevefelt1975rates, zygelman2003}
according to:

\begin{subequations}
\label{rateequation}
\begin{align}
  \frac{\mathrm d\rho_n}{\mathrm dt} = &
    \, \rho_{\rm e}(t) \, \sum_{n^{\prime} \ne n} \left[
\rho_{n^{\prime}}(t) \,  R(n^{\prime},n) -  \rho_n(t) R(n,n^{\prime})  \right]\notag \\
    &   + \rho_{\rm e}(t)^3 R_{\rm rec}(n)-\rho_{\rm e}(t) \, \rho_n(t) R_{\rm ion}(n),
  \label{rateequation1} \\
  \frac{\mathrm d\rho_{\rm e}}{\mathrm dt} = &
   \,\rho_{\rm e}(t) \, \sum_{n^{\prime}=1} \left[\rho_{n^{\prime}}(t)
R_{\rm ion}(n^{\prime}) - \rho_{\rm e}(t)^2 R_{\rm rec}(n^{\prime})\right] .
  \label{rateequation2}
\end{align}
\end{subequations}
We use the transition rates recently determined in \citep{pohl2008rates} by CTMC
calculations.
The rate for excitation from the atomic level $n$ to level $n^{\prime}$ is given by
\begin{equation}\label{coll1}
  R(n,n^{\prime}) = k_0 \, \epsilon_{n^{\prime}}^{3/2}  \, e^{\epsilon_{n^{\prime}} -
\epsilon_n}
          \left[ \frac{22}{(\epsilon_n + 0.9)^{7/3}} +
                 \frac{9/2}{\epsilon_n^{5/2} \, \Delta \epsilon^{4/3}}
          \right],
\end{equation}
the rate for de-excitation from $n$ to $n^{\prime}$ is given by
\begin{equation}\label{coll2}
  R(n,n^{\prime}) = k_0 \, \frac{\epsilon_n^{5/2}}{\epsilon_{n^{\prime}}} \,
          \left[ \frac{22}{(\epsilon_{n^{\prime}} + 0.9)^{7/3}} +
                 \frac{9/2}{\epsilon_{n^{\prime}}^{5/2} \, \Delta \epsilon^{4/3}}
          \right],
\end{equation}
the rate for recombination into an atomic level $n$ is given by
\begin{equation}\label{coll3}
  R_{\rm rec}(n)=
    \frac{11 \, \sqrt{\mathcal{R} / k_\text{B} T} \, k_0 \, n^2 \,\Lambda^3}
        {\epsilon_n^{7/3} + 4.38 \, \epsilon_n^{1.72} + 1.32 \epsilon_n},
\end{equation}
and the rate for ionization of atoms in level $n$ is
\begin{equation}\label{coll4}
  R_{\rm ion}(n) = \frac{11 \, \sqrt{\mathcal{R} / k_\text{B} T} \, k_0 \,
e^{-\epsilon_n}}
                {\epsilon_n^{7/3} + 4.38 \, \epsilon_n^{1.72} + 1.32 \epsilon_n},
\end{equation}
where
$k_0 = e^4 / (k_\text{B}T \sqrt{m \, \mathcal{R}})$,
$\epsilon_n = \mathcal{R} / (n^2 \, k_\text{B} T)$,
$\Delta \epsilon = |\epsilon_n - \epsilon_{n^{\prime}}|$,
$\Lambda = \sqrt{h^2 / (2 \pi \, m \, k_\text{B}T)}$ is the thermal de Broglie wavelength,
$\mathcal{R}$ is the Rydberg constant,
and $h$ is Planck's constant.

\section{Kinetic Bottleneck}
\label{sec_bottleneck}
Generally, the recombination rate $\nu$ is defined as the rate at which ground state atoms
are populated
in the plasma \citep{bates1962rates, stevefelt1975rates, zygelman2003, zygelman2005}.
While such deeply bound states defy a classical description, it was shown in
\citep{mansbach1969} that this rate can also be determined from the downward energy flux
through a kinetic bottleneck energy that divides weakly bound from stable atomic states.
As the bottleneck typically lies in the classical region of binding energies
this process can be described classically.

The concept of the kinetic bottleneck is readily understood from the following simple
arguments.
Depending on its binding energy $E_\text{bn}$, a bound electron has a certain probability
$P_{\text{ion}}(E_\text{b})$ for collisional re-ionization and a probability to be
successively driven to deeper binding energies until it eventually reaches the ground
state without being re-ionized on its way. 
In our simulations the latter equals the probability $P_{\text{sink}}(E_\text{b}) =1 -
P_{\text{ion}}(E_\text{b})$  for reaching the energy sink at $E_{\rm sink}$, since
$P_{\text{ion}}(E_{\rm sink})\approx0$.
The re-ionization probability decreases with deeper binding and ultimately falls below the
recombination 
probability, such that the atomic states become more and more stable against ionizing
electron-atom collisions.
Hence, the kinetic bottleneck is defined as the energy $E_{\rm bn}$ at which recombination
starts to dominate, i.e. 
the energy at which 
\begin{equation}
  P_{\text{ion}}(E_\text{b}=E_{\rm bn}) =  P_{\text{sink}}(E_\text{b}=E_{\rm bn}) =
\frac{1}{2}.
  \label{ionprob_bn}
\end{equation}
Three-body CTMC calculations predict a simple linear scaling of the bottleneck
\citep{mansbach1969}
\begin{equation}
 E_{\rm bn} \approx - 3.83 \, \Gamma^{-1} \frac{e^2}{a}.
 \label{bottleneck}
\end{equation}

\begin{figure}[t!]
 \includegraphics[width=\columnwidth]{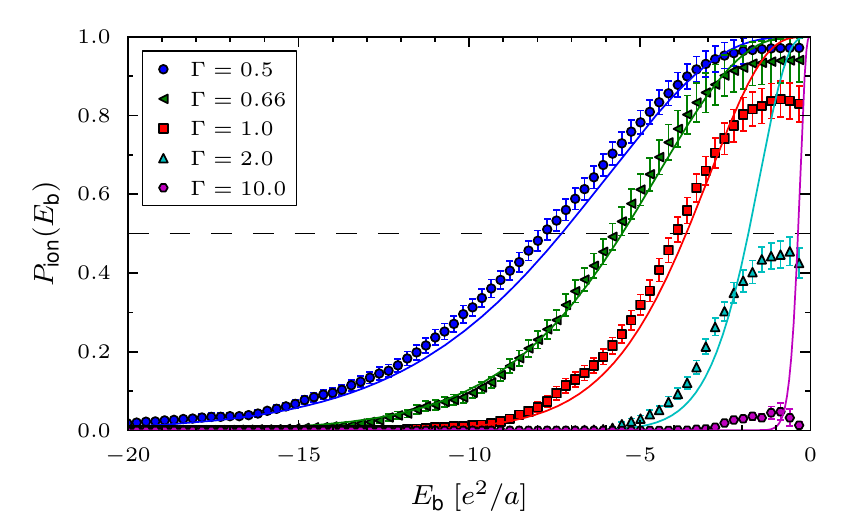}
 \caption{(color online) Ionization probability $P_{\text{ion}}(E_\text{b})$ for different
          $\Gamma$. Symbols correspond to MD-data with $\phi_{\rm c} = 4 \, \pi$.
          The lines show the ideal plasma prediction obtained by eq. \eqref{rateion}. The
        crossing with $P_\text{ion} = \frac{1}{2}$ (dashed line) determines the location
        of the kinetic bottleneck $E_{\rm bn}$.}
 \label{ionprob}
\end{figure}

As the bottleneck energy is crucial for determining the recombination rate, we first need
to check the validity of this simple law in the strong coupling regime.
In the MD simulations, $E_{\rm bn}$ can also be determined from  eq. \eqref{ionprob_bn},
where
$P_{\text{ion}}(E_\text{b})$ is calculated from the corresponding bound state energy
densities obtained from the above described
energy trajectories (see Fig. \ref{energy}) according to
\begin{equation}
 \varrho_{\text{tot}}(\epsilon) = \langle \int_{0}^{\tau} \delta (E_\text{b}(t) -
\epsilon) \,
\text{d}t \
\rangle,
 \label{energy_density}
\end{equation}
where $\tau$ is the simulation time of a single simulation run and $\langle \dots \rangle$
denotes the average over the statistical ensemble.
This total energy density can be split into two parts, 
$\varrho_{\text{tot}}(E_\text{b}) =
\varrho_{\text{ion}}(E_\text{b}) + \varrho_{\text{rec}}(E_\text{b})$.
$\varrho_{\text{ion}}(E_\text{b})$ contains only bound states that are subsequently
ionized, while $\varrho_{\text{rec}}(E_\text{b})$ counts only energies of bound
states that reach the energy sink without intermediate re-ionization (red and blue,
respectively, in Fig. \ref{energy}b).
The ionization probability $P_{\text{ion}}(E_\text{b})$ is then obtained from the ratio
\mbox{$P_{\text{ion}}(E_\text{b}) =
\frac{\varrho_{\text{ion}}(E_\text{b})}{\varrho_{\text{tot}}(E_\text{b})}$}, and shown in
Fig. \ref{ionprob} for several values
of the Coulomb coupling parameter $\Gamma$.

For ideal plasmas ($\Gamma\rightarrow0$) and within the adiabatic treatment of Bates,
Kingston, and McWhirter \citep{bates1962rates}, the ionization
probability $P_{\rm ion}(E_\text{b} = -\mathcal{R} / n^2)$ can be directly obtained from
the collision rates eqs. (\ref{coll1})-(\ref{coll4}) \citep{burgess1976rateequations,
zygelman2005}
\begin{equation}
  P_\text{ion}(E_\text{b}) =
      \frac{R_{\rm ion}(n)}{A(n)} \, + 
    \sum_{n^{\prime}=1} \frac{R(n,n^{\prime}) \, R_{\rm ion}(n^{\prime})}
        {A(n) \, A(n^{\prime})} + 
    \dots,
  \label{rateion}
\end{equation}
by summing over the probabilities of all possible pathways in energy space that connect an
atomic level of binding energy $E_\text{b} = -\mathcal{R} / n^2$ to the
continuum, where $A(n) = \sum_{n^{\prime} \ne n}^{n_\text{max}} R(n,n^{\prime}) + R_{\rm
ion}(n)$ is the total rate
for leaving level $n$. The first term in eq. \eqref{rateion} represents the probability
that
the bound electron will be ionized directly from level $n$. The second term accounts for
an intermediate step via a level $n^{\prime}$ from which subsequent ionization occurs, and
so
on.

The good agreement between our MD results and eq. (\ref{rateion}) in the regime of small
to moderate $\Gamma$, shown in Fig. \ref{ionprob}, attests to the accuracy of both
approaches. In this regime the bottleneck energy can be straightforwardly determined
according to eq. \eqref{ionprob_bn} and corresponds to the intersections of
$P_\text{ion}$ and the horizontal dashed line at 0.5 in Fig. \ref{ionprob}. At larger
$\Gamma$-values, however, the ionization probability is suppressed to
$P_\text{ion}(E_\text{b}) < 0.5$ over the entire range of binding energies, such that the
bottleneck energy vanishes.

\begin{figure}[b!]
 \includegraphics[width=\columnwidth]{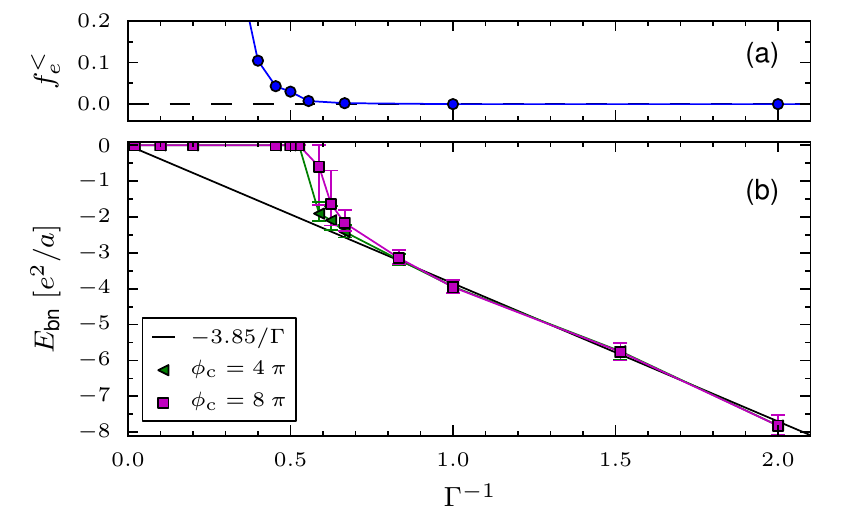}
 \caption{(color online) (a) Fraction $f_e ^<$ of free electrons with energy \mbox{$E <
          -4\, \Gamma^{-1} \, \frac{e^2}{a}$}.
          (b) Bottleneck energy $E_\text{bn}$ as a function of $\Gamma$ for two different
          critical angles $\phi_{\rm c}$ compared to the standard \mbox{$-3.85 \,
            \Gamma^{-1} \, \frac{e^2}{a}$}
           scaling (black line).
          }
 \label{fig_bottleneck}
\end{figure}

The resulting temperature dependence of the bottleneck energy is shown in
Fig. \ref{fig_bottleneck}b.
For small coupling parameters the MD simulations predict a linear scaling, $E_{\rm
bn} = -3.85 \, \Gamma^{-1} \frac{e^2}{a}$,  in quantitative agreement
with the three-body CTMC result, eq. (\ref{bottleneck}). However, 
for $\Gamma \gtrsim 2$, the bottleneck drops to zero. In contrast to the ideal plasma
case, where electrons 
still have to overcome the kinetic bottleneck barrier before recombination, stable atoms
are formed directly 
in the strongly coupled regime. The disappearance of the kinetic bottleneck can be traced
back to correlation-induced 
continuum lowering, which around $\Gamma\approx2$ leads to a merging of the ionization
threshold and the bottleneck 
energy. To demonstrate this point, Fig. \ref{fig_bottleneck}a shows the fraction $f_{\rm
e}^<$ of \emph{free} plasma electrons
with a total energy of $E_i < -4 \, \Gamma^{-1} \, \frac{e^2}{a}$. The simulation results
yield a steep increase of $f_{\rm e}^<$ around 
$\Gamma \approx 2$, at which the bottleneck, thus, has to disappear, in agreement with
Fig. \ref{fig_bottleneck}b.
Consequently, the critical angle $\phi_{\rm c}$ has almost no effect on the critical
$\Gamma$ at which the bottleneck energy drops to zero and can only slightly affect the
value of $E_{\rm bn}$ for smaller $\Gamma$ (see Fig. \ref{fig_bottleneck}b).

\section{Recombination Rate}
\label{sec_rate}
\begin{figure}[t!]
 \includegraphics[width=\columnwidth]{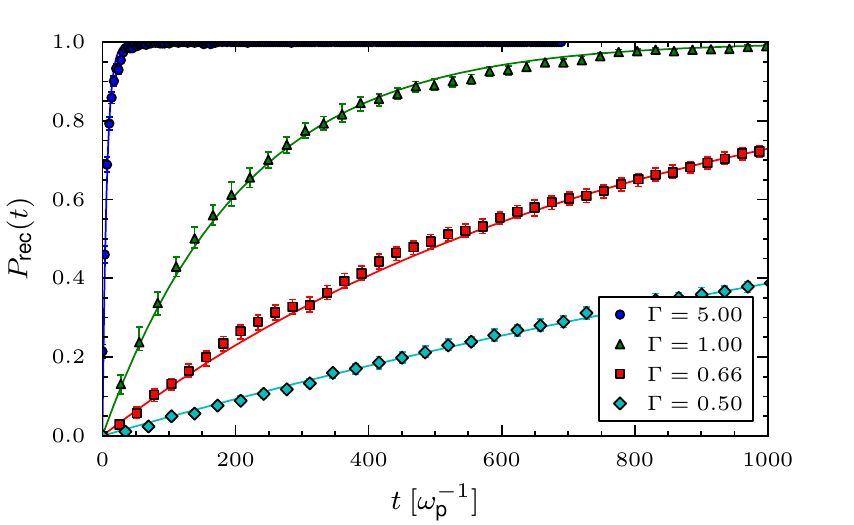}
 \caption{(color online) Recombination probability $P_{\text{rec}}(t)$ as a function of
          time for different coupling strength $\Gamma$ and $\phi_{\rm c} = 4 \pi$.
          The lines represent fits of eq. \eqref{rec_proba} to the MD
simulation result.}
 \label{recproba}
\end{figure}

Having determined the location of the bottleneck we can now proceed to extract the
recombination rate $\nu$ from our MD simulations. 
This is done in a straightforward manner by calculating the time dependent recombination
probability $P_{\text{rec}}(t)$, defined as the
probability to observe a bound electron with binding energy $E_\text{b}(t)<E_{\rm bn}$ at
a time $t$.
Fig. \ref{recproba} shows examples of the obtained $P_{\text{rec}}(t)$ for
different coupling strength and a critical angle $\phi_{\rm c} = 4 \pi$.
The numerical data is well fitted by an exponential bound state relaxation law of the form
\begin{equation}
  P_{\text{rec}}(t) = 1 - e^{-\nu t},
  \label{rec_proba}
\end{equation}
which permits to extract the recombination rate $\nu$.

\begin{figure}[t!]
  \includegraphics[width=\columnwidth]{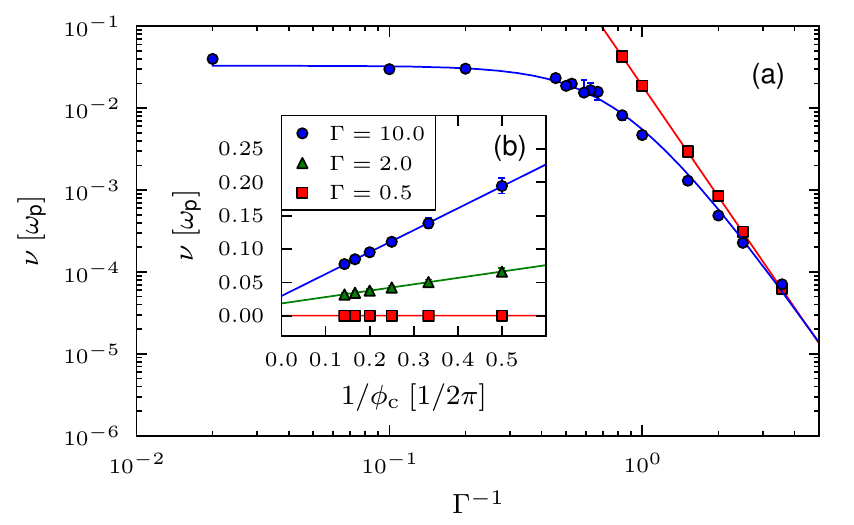}
  \caption{(color online) (a) Recombination rate $\nu$ as a function of inverse coupling
          strength $\Gamma^{-1}$ calculated with MD simulations (circles)
          compared to the $\Gamma^{-9/2}$-scaling obtained by eq. \eqref{recrate_re}.
          The red line corresponds to eq. \eqref{tbr_coefficient}, the blue line serves as
          guide to the eye.
          (b) Dependence of $\nu$ on inverse critical angle $1/\phi_{\rm
          c}$ with fitted extrapolation (lines).
          }
  \label{rate}
\end{figure}

Fig. \ref{rate}b shows the rate $\nu$ as a function of the inverse critical angle
$\phi_{\rm c}$.
One finds a linear dependence on $1/\phi_{\rm c}$, whose slope tends to increase with
increasing coupling strength. The fact that in the considered range of $\phi_{\rm c}$ all
simulation results perfectly lie on a line, allows us to extrapolate to
$1/\phi_{\rm c}\rightarrow0$, corresponding to stable atomic states.

For comparison with the weak coupling CTMC results, we also calculate the recombination
rate \cite{zygelman2005}
\begin{equation}
  \nu = \sum_{n} \left[1 - P_\text{ion}(-\mathcal{R}/n^2) \right] \, R_{\rm rec}(n)\;,
  \label{recrate_re}
\end{equation}
as obtained from eq.(\ref{coll3}) and (\ref{rateion}). Fig. \ref{rate}a shows the
extrapolated many-body MD (circles) and 
three-body CMTC (squares) results for the recombination rate as a function of $\Gamma$. In
the weak coupling regime 
we find good quantitative agreement with the $T^{-9/2}$ scaling \cite{pohl2008rates}
\begin{equation}
  \nu \approx 0.019 \, \omega_\text{p} \, \Gamma^{9/2} \;.
  \label{tbr_coefficient}
\end{equation}

Notably, the MD results demonstrate the high accuracy of the rate equation description
even for moderate coupling 
strength $\Gamma\lesssim0.3$, corresponding to typical parameters in the long-time
evolution of UCPs 
\cite{roh02,fzl07}. In the regime of strong Coulomb coupling, however, one finds a
significant suppression of the 
recombination rate. Our result approaches a constant value of $\nu\sim0.03\omega_{\rm p}$
with increasing 
$\Gamma$, thereby resolving the apparent timescale paradox described in section
\ref{sec_introduction}.
 At intermediate $\Gamma$-values, our results are consistent with previous MD simulations
of two-component 
 plasmas \cite{kuzmin2002numsimplasma}. These two-component simulations also predict a
suppression by 
 a factor of $\sim2$ for the particular value of $\Gamma=0.6$ studied in
\cite{kuzmin2002numsimplasma}, 
 suggesting that the present OCP model should provide a good description
of recombination 
 in neutral plasmas. In this case, however, additional disorder-induced electron heating
\cite{mck02,kun02} due to the strong 
 attractive electron-ion interaction limits the range of realizable Coulomb coupling
parameters, as will be briefly 
 discussed below.
 
\section{Two-component plasma simulations}
\label{sec_full}
We also performed MD simulations of a two component plasma with $N$
ions and $N$ electrons in a cubic simulation cell with PBC. For these simulations, all
interactions 
and the corresponding PBC are calculated within the FMM. We use a very small global time
step 
$\Delta t=10^{-5}$, ensuring an accurate treatment of even the lowest bound states
observed in the simulations.
In analogy to the previously described simulation scheme, the full plasma simulations
start with $N$ randomly distributed atoms which are photoionized at $t=0$ as detailed in
section 
\ref{sec_md}. The initial kinetic excess energy $E=\frac{3}{2\Gamma_0}\frac{e^2}{a}$
determines the 
initial effective coupling strength $\Gamma_0$, i.e. the scaled kinetic energy
$\Gamma_0^{-1}$. 

\begin{figure}[t!]
 \includegraphics[width=\columnwidth]{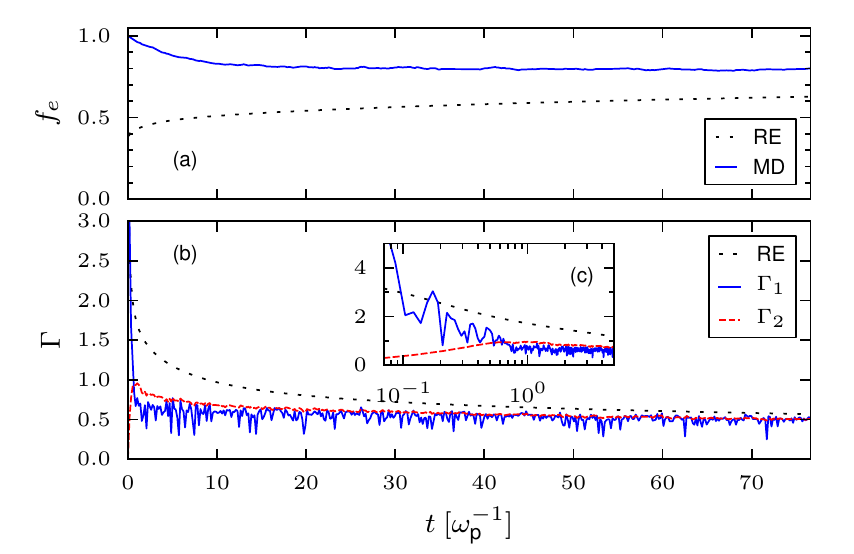}
 \caption{(color online) Evolution of the normalized free electron density $f_e$ (a) and
          electronic coupling parameter $\Gamma$ (b) calculated with MD simulation
          for $\phi_{\rm c} = 4 \, \pi$ and rate equation (RE) for an initial excess
          energy corresponding to coupling strength of $\Gamma_0 = 50$.}
 \label{fullsim_high}
\end{figure}

\begin{figure}[t!]
 \includegraphics[width=\columnwidth]{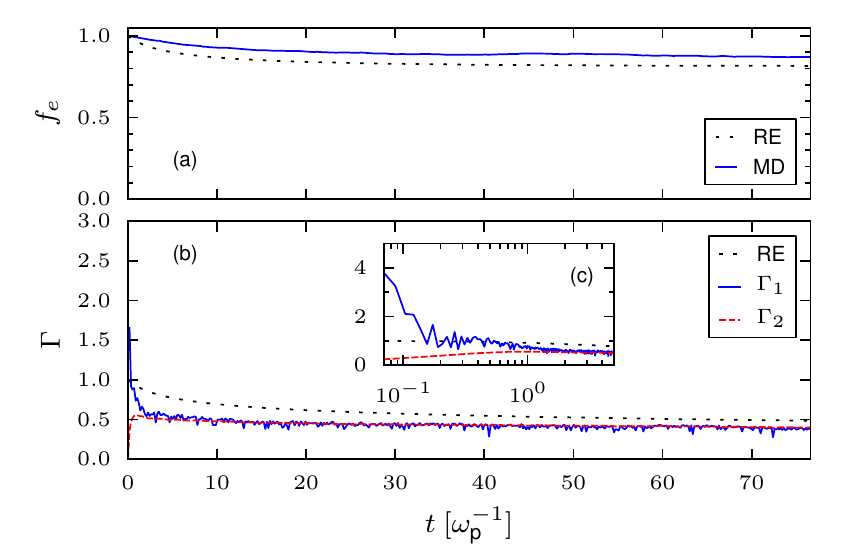}
 \caption{(color online) Same plot as in Fig. \ref{fullsim_high} but for $\Gamma_0 = 1$.}
 \label{fullsim_low}
\end{figure}

However, since this procedure creates a highly non-equilibrium plasma, it takes a finite
time to establish 
a well defined electron temperature. In order to characterize the corresponding initial
relaxation we monitor the evolution of
two different coupling parameters, defined through $i$th order momenta $\langle v^i
\rangle$ of the free-electron 
velocity distribution
\begin{subequations}
\label{gamma}
\begin{align}
  \Gamma_1 & = 3 / \langle v^2 \rangle \\
  \Gamma_2 & = \sqrt{6 / (\langle v^4 \rangle - \langle v^2 \rangle^2)}.
\end{align}
\end{subequations}
In local equilibrium, i.e. once the electrons have established a Maxwellian velocity
distribution, 
$\Gamma_1 = \Gamma_2$. Indeed the simulation results shown in Figs. \ref{fullsim_high}
and 
\ref{fullsim_low} show that both definitions of $\Gamma$ approach each other on a
timescale 
$\omega_{\rm p}^{-1}$, such that one can speak of an electron temperature for 
$t>\omega_{\rm p}^{-1}$. 

On the same time scale, the electrons heat up due to disorder-induced heating 
\cite{mck02,kun02}. As shown in Figs. \ref{fullsim_high} and \ref{fullsim_low}, for both
very large 
($\Gamma_0=50$, Fig.\ref{fullsim_high}) and moderate  ($\Gamma_0=1$,
Fig.\ref{fullsim_low}) 
initial effective coupling strength, the Coulomb coupling parameter relaxes to a value of 
$\Gamma\approx0.5$ during the initial relaxation stage, which is a factor of $\sim2$
smaller than 
found in \cite{kun02} but agrees with the findings of more recent MD simulations
\cite{ngr11}.

We have investigated the amount of initial heating for a range of initial conditions, 
including highly pre-ordered states, where ions and electrons have been placed on  
regular lattice structures. In contrast to the ionic plasma component, where such a
pre-ordering leads 
to significant suppression of the heating due to the repulsive ion-ion interactions
\cite{gem03,ppr04c}, 
the attractive electron-ion interaction is found to cause electron heating to
$\Gamma\lesssim0.5$ 
irrespective of the initial state. This clearly excludes the existence of metastable, very
strongly coupled 
two-component plasma states, in which recombination is suppressed by orders of magnitude,
as has been 
suggested recently \cite{norman2009recombination,bbz11} on the basis of numerical
simulations.
 
Nevertheless, the results of the previous sections have shown that the recombination
dynamics differs
significantly already for $\Gamma\approx0.5$, yielding a suppression of $\nu$ by a factor
of $~2$. A 
comparison to the time evolution of the free electron number obtained from the rate
equations 
(see section \ref{sec_rateequations}) reveals significant deviations, which increase with
decreasing 
initial energy (see Figs. \ref{fullsim_low} and \ref{fullsim_high}). Moreover, despite the
fact that $\Gamma$ 
has relaxed to $\sim0.5$ after $\sim50 \, \omega_{\rm p}^{-1}$ for both $\Gamma_0=1$ and
$\Gamma_0=50$,
the number of weakly bound Rydberg atoms differs by about a factor of $\sim2$, which we
attribute to the 
longer relaxation time of bound states. We anticipate, that these effects and deviations
from the traditional 
treatment of recombination may be observable via short-time probing of UCP dynamics.

\begin{figure}[b!]
  \includegraphics[width=\columnwidth]{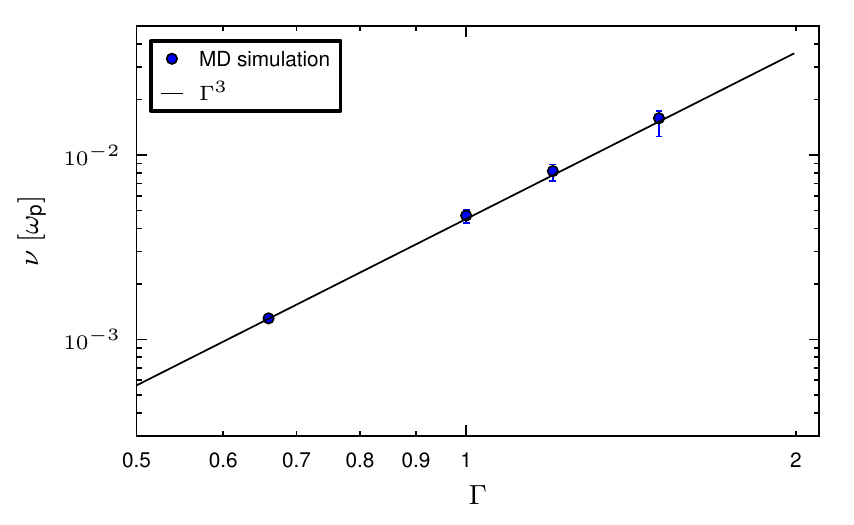}
  \caption{(color online) Recombination rate $\nu$ calculated with MD simulation
          (circles) for intermediate coupling strength $\Gamma$. At the onset of
          strong correlations the rate scales with $\Gamma^3$ (line).
          }
  \label{rate_crossover}
\end{figure}

Indeed, recent measurements of recombination fluorescence on a sub-microsecond time scale
suggest such deviations \cite{br08}. In this experiments, the time dependent fluorescence from
low-lying transitions 
of recombined atoms has been measured with a time resolution $<100$ ns. At high temperatures,
i.e. in the weakly 
coupled regime, the initial signal $S(t)$ was found to rise proportional to $\rho_{\rm
e}^3$, consistent with the 
picture of isolated three-body collisions, for which 
$S(t)\sim\rho\nu t \propto \rho \omega_{\rm p}t\Gamma^{9/2} \propto \rho_{\rm e}^3$. For
lower temperatures, detailed 
simulations based on three-body CTMC rates predict a density scaling $\sim \rho_{\rm
e}^{1.8}$, while the 
experiment shows a scaling $\sim \rho_{\rm e}^{2.2}$. Here it is interesting to note, that
the recombination rate around $\Gamma \approx 1$ already shows a different scaling
$\nu\sim\omega_{\rm p}\Gamma^3$ (see Fig.\ref{rate}), giving $\rho\nu t\propto \rho_{\rm e}^{2.5}$. As
described in \cite{br08}, the 
fluorescence signal is, however, determined also by the initial disorder-induced heating
as well as heating due 
to the formation of Rydberg atoms themselves, such that this simple comparison should be
regarded as qualitative only.
Nevertheless, MD simulations, as described in this work, combined with a detailed treatment of the radiative
cascade of deeply bound states to compare with such measurements may
elucidate the role of correlation effects in recombination dynamics of UCPs.

\section{Summary}
\label{sec_conclusion}
We have presented extensive numerical simulations of Rydberg atom formation
in 
 plasmas, that take into account correlations and many-body interactions between the
plasma electrons.
This allows to stretch the focus of such studies deep into the strongly coupled regime,
beyond the range of validity of three-body CTMC calculations.

We find quantitative agreement for the recombination rate with previous rate equation
calculations 
\cite{pohl2008rates} in the weakly coupled regime. Such simplified treatments are shown to
yield an
excellent description, even for Coulomb coupling strengths of up to $\Gamma=0.3$, which
covers the typical $\Gamma$-values obtained in the long-time dynamics of UCPs \cite{roh02,fzl07}.

However, as the electron plasma becomes strongly coupled the bottleneck is found to 
disappear in the lowered continuum due to increasing electron-electron correlations.
In this strongly coupled regime, $\nu$ is shown to approach a constant value well below the
plasma frequency $\omega_\text{p}$, resolving the temperature-divergence problem of the 
common three-body recombination rate, $\nu\sim T^{-9/2}$, in the ultracold domain.

MD simulations of two component plasmas show that the achievable coupling strength in
neutral plasmas is limited to $\Gamma\approx0.5$, in agreement with recent simulations
discussed  in \cite{ngr11} while contradicting the findings of
\cite{norman2009recombination,bbz11}.
Nevertheless, a comparison to the MD results for such coupling parameters suggest that 
deviations from common rate equation descriptions may be observable in the short-time 
dynamics of UCPs. We finally note, that recent experiments on molecular ultracold plasmas 
\cite{grant2008supersonic,grant2009nobeam}, realizing much higher densities than atomic 
systems, show strong deviations from the expansion behavior of atomic systems \cite{kkb00}, 
which, thus far, has been well described within simple rate equation treatment of Rydberg atom 
formation \cite{roh02,lgs07,gls07}. Exploring the origin of these deviations, however, 
requires to account for additional molecular processes \cite{sms11}, that may also alter the 
plasma expansion behavior.

\begin{acknowledgments}
We thank U. Saalmann and F. Robicheaux for valuable discussions and comments, and are 
grateful to I. Kabadshow for support with the FMM.
\end{acknowledgments}
\bibliography{./tbr_bibo}
\end{document}